\documentclass{aastex}
\begin{document}
\title{Discovery of the Orbit of the Transient X ray Pulsar SAX J2103.5+4545 
 }

   \author{Altan Baykal $^{1}$, 
          Michael J. Stark $^{2}$, 
         Jean Swank $^{3}$ }  
\affil{ 
 $^{1}$ Physics Department, Middle East Technical University,
  Ankara 06531, Turkey; 
 altan@astroa.physics.metu.edu.tr\\
$^{2}$ Department of Physics and Astronomy Denison University Granville, 
OH 43023, USA;
starkm@marietta.edu\\ 
 $^{3}$ Laboratory for High Energy Astrophysics NASA/GSFC
  Greenbelt, Maryland 20771 USA; 
 swank@pcasun1.gsfc.nasa.gov }

   \begin{abstract}
Using X-ray data from the Rossi X-Ray Timing Explorer (RXTE), 
we carried out pulse timing analysis of the transient X-ray pulsar
SAX J2103.5+4545. 
An outburst was detected  by All Sky Monitor (ASM) October 25 1999
and reached a peak X-ray brightness of 27 mCrab October 28.
 Between November 19 and December 27, the RXTE/PCA carried out pointed
 observations which provided us with pulse arrival times.
These yield an 
eccentric orbit (e=$ 0.4 \pm 0.2$) with an orbital period of 
$12.68 \pm 0.25$ days and light travel time across
the projected semimajor axis of $ 72 \pm 6$ sec.
The pulse period was measured to be $358.62171 \pm 0.00088 $ s
and the spin-up rate $(2.50 \pm 0.15) \times 10^{-13}$ Hz s$^{-1}$.
The ASM data for the February to September 1997 outburst in which 
BeppoSAX discovered 
SAX J2103.5+4545
(Hulleman, in't Zand and Heise 1998)
are modulated at 
time scales close to the orbital period. Folded light curves of the 1997 ASM 
data and the 1999
PCA data are similar and show that 
the intensity increases at periastron passages.

\ Subject headings: {binaries: close -- pulsars: individual: SAX J2103.5+4545 --
stars: neutron -- X-rays: stars 
} 

\end{abstract} 
\section{Introduction}
About 50 accretion powered X-ray pulsars are known. 
Although most of them are in binary systems, 
binary orbital parameters are known for only half of them
(Bildsten et al. 1997). 
The optically identified companions are generally high mass stars 
(van Paradijs 1995). 
Among the high mass companions half of them are giant or supergiant 
stars and the other half are Oe and Be stars with episodic mass loss events. 

 The X-ray source SAX J2103.5+4545 was discovered using the BeppoSAX X-ray 
 satellite  (Hulleman, in't Zand, and Heise 1998).
 Pulsations at 358.61 s were found. The X-ray spectrum 
 was a power law with photon number index of 1.27, typical of accreting 
high--field
 pulsars. In analogy to other X-ray pulsars 
 this object was proposed to be a neutron star in a binary.  
It was noted that the  B8  star HD 200709 
was at the edge of the BeppoSAX error box. 

 In this work, we report on  
 X-ray observations made with the Rossi X-ray Timing
 Explorer (RXTE) during an outburst two years later. 
We  determine the 
 orbital parameters using pulse arrival time analysis (see also  
 Baykal, Stark, and Swank 2000).

\section{Observations} 

 SAX J2103.5+4545 was observed during 1999 November 19 and December 27 
for an RXTE guest observer proposal to study the decays of transient 
pulsars. The results presented here are 
based on data collected with the Proportional Counter Array (PCA)
(Jahoda et al. 1996) and All Sky Monitor (ASM) (Levine et al. 1996).
The PCA instrument consists of five identical multi-anode proportional
counter units (PCUs). The PCA operates in the 2-60 keV energy range,
with a total effective area of approximately 7000 cm$^{2}$ and a field of
view $\sim 1^{\circ}$ FWHM. The ASM consists of three identical
Scanning Shadow Cameras (SSCs) each of
 which has a net active area for detecting
X-rays of $\sim $30 cm$^{2}$. 
The ASM  sensitive range is 2 to 12 keV. It perform sets of 90 sec
pointed observations, covering $\sim$80$\%$ of the sky every  
$\sim$ 90 minutes.  

\section{Results}
Background light curves were generated by using the background estimator 
models based upon the rate of very large events, spacecraft activation, 
and cosmic X-ray emission with the standard PCA analysis tools. 
The background light curves were subtracted from the source light curves 
which were obtained from the Good Xenon event data. 
The background subtracted light curve is presented in figure 1.
The observation times were corrected with respect to the barycenter of
the Solar system. Then summed power spectra were used to estimate the 
average pulse frequency. Pulse profiles were obtained by folding the 
lightcurves at the pulse period deduced from the power spectra.
A template pulse profile is presented in figure 2.
Pulse arrival times were found by cross-correlating the pulse 
profiles. In the pulse timing analysis, we have used the harmonic 
representation of pulse profiles (Deeter $\&$ Boynton 1985). 
The pulse profiles were expressed in terms of harmonic series and 
cross-correlated with the template pulse profile. 
Figure 3, presents the pulse arrival times. The pulse arrival times 
showed a quadratic trend superposed on orbital Doppler delays. 
The pulse arrival times can be written as (Deeter, Boynton, $\&$ Pravdo 1981)

\begin{equation}
\delta t = \frac{\delta P}{P} (t-t_{0})
+\frac{1}{2}\frac{\dot P}{P}(t-t_{0})^{2}+x\sin(l)
-\frac{3}{2}xe\sin(w)+\frac{1}{2}xe\cos(w)\sin(2l)
-\frac{1}{2}xe\sin(w)\cos(2l).
\end{equation} 
Here t$_{0}$ is the mid-time of the observation; 
 $\delta P$ is the deviation from mean pulse period $P$; 
 $\dot P$ is the time derivative of the pulse period; 
$x=a_{x} \sin(i)/c$ is 
 the light traveltime 
 for projected semimajor axis 
(where i is the inclination angle between the line of sight 
and the orbital angular momentum vector); 
$l=2\pi (t-T_{\pi/2})/P_{orbit}+\pi/2$ is the mean orbital longitude 
at time $t$;
 $T_{\pi/2}$ is the epoch when the mean orbital longitude is equal to 
90 $^{\circ}$; 
 $P_{orbit}$ is the orbital period; $e$ is the eccentricity; and $w$ is the 
 longitude of periastron. The above expression is fitted to the pulse 
arrival times data. Table 1 presents the timing solution of 
SAX J2103.5+4545. Figure 4 presents the pulse arrival times after the
 removal of the quadratic trend (or intrinsic -$\dot P$). 
 The source was spinning up 
during the observation at the rate of
 $\dot \nu =(2.5 \pm 0.15) \times 10^{-13}$ Hz s$^{-1}$. 
The periodic trend of the pulse arrival times yields 
an eccentric orbit (e=$ 0.4 \pm 0.2$) with a orbital period of
$12.68 \pm 0.25$ days.

The PCA lightcurve is folded at the orbital period of 12.68 days
with respect to the reference time of mean longitude 90$^{\circ}$.
The folded PCA lightcurve is presented in figure 5.
 In this figure, orbital phase $\sim $0.4 corresponds to the
periastron passage of the neutron star. 
It is clearly seen that the X-ray flux (or count rate) is increased
at the periastron passage of the X-ray pulsar.
This is strongly suggesting that the mass accretion rate onto the
neutron star is increased during the periastron passage.

Since 
SAX J2103.5+4545 was observed with the BeppoSAX satellite from February to 
September 1997 (Hulleman, in't Zand and Heise 1998)  
we extracted the public ASM data and presented 3 days average of ASM light 
curve in figure 6. 
The X-ray pulsar was actively seen in ASM in 1997 for $\sim$ 200 days. 
 In order to see 
the X-ray flux modulations due to the eccentric orbit, we 
employed a $\chi ^{2}$ search. 
The result is shown in figure 7;  
there is obvious peak centered at 12.6 days. 
The  ASM data folded on the orbital period 
is shown in  figure 8. The folded ASM light curve also shows an increase 
in X-ray flux at the periastron passage. The ASM and PCA lightcurves 
are quite similar. 

The derived orbital parameters yield a mass function for the  
companion 
\begin{equation}
f(M)=\frac{4\pi ^{2}(a_{x}\sin i)^{3}}{GP^{2}_{orb}}
=\frac{(M_{c} sin i )^{3}}{(M_{x}+M_{c})^{2}}
\sim 2.5 M_{\odot}.
\end{equation}
For an X-ray pulsar mass of ($\sim$1.4M$_{\odot}$) and an inclination of 
45$^{\circ}$, the companion mass would be  
$\sim $7$M_{\odot}$.

\section{Discussion}

High mass X-ray binaries fall into three separate groups when 
the pulse periods are compared with the orbital periods (Corbet 1986). 
The systems with Be companions show correlations between orbital period 
and spin periods (Corbet 1986, van Kerkwijk 1989), while systems with   
giant and supergiant companions fall into two separate regions. 
X-ray pulsars with Be type companions show transient behavior and
these pulsars generally accrete from
the variable envelope of a Be star.
The mass accretion increases at the periastron passages since the 
density of plasma is greater when the pulsar is close to the companion star. 
If SAX J2103.5+4545 has a Be type companion, according to the 
Corbet diagram, a pulse period of $\sim$ 358 s implies
an orbital period of $\sim$190 d (Hulleman, in't Zand and Heise 1998).
The measured orbital period and pulse period of SAX J2103.5+4545 
suggests that the companion star has to be a giant or a supergiant star
(Corbet 1986). It is interesting to note that even if this pulsar
has a supergiant companion it is still a transient X-ray pulsar.  


Hullemann, in't Zand, \& Heise (1998) attempted to use the theoretical
relation between $\dot P$ and X-ray luminosity to test the consistency of 
possible locations of the source, but their upper limits on $\dot P$ were
not restrictive. According to Ghosh \& Lamb (1979) the maximum spin-up rate
($-\dot P$) is given by

 \begin{equation}
- \dot P =  2.2\times 10^{-12} \mu_{30}^{2/7} m_{x}^{-3/7} 
R_{6}^{6/7} I_{45}^{-1} P^{2} L_{37}^{6/7}~~s~~ s^{-1}
\end{equation} 
where $\mu_{30}$, $m_{x}$, $R_{6}$, and $I_{45}$ are the magnetic dipole 
moment in units of $10^{30}$ G cm$^{3}$, the mass in units of solar mass 
($m_{x}=M_{x}/M_{\odot}$), the radius in units of 10$^{6}$ cm
 and the moment of inertia 
in units of 10$^{45}$ g cm$^{2}$, respectively, and 
$L_{37}$ represents the X-ray luminosity in units of $10^{37}$ erg s$^{-1}$.
The spin-up rate we observed ($-3.2 \times 10^{-8}$ s s$^{-1}$ )implies
an accretion luminosity of
$ L_{x} = GM\dot M /R \sim 8 \times 10^{35}$ erg s$^{-1}$ 
for a typical surface magnetic field strenght of $\sim 10^{12}$ Gauss.

The average flux seen by the ASM
during the period of our PCA measurements was only 5 mCrab and the PCA
average spectrum implied a total unabsorbed flux 
of $2.4 \times 10^{-10}$ ergs cm$^{-2}$ s$^{-1}$. 
A distance can be estimated from this flux and the X-ray luminosity 
estimate given above; the result is   
D = $\sqrt{\frac{L_{x}}{4\pi F_{x}}} \sim 5$ 
 kpc. 
is about 5 kpc.
The B star HD 200709 was suggested as a marginal candidate of optical 
counterpart at a distance of 
$\sim 0.7$ kpc (Hulleman, in't Zand and Heise 1998).   
Our finding suggests that SAX J2103.5+4545 is far beyond the star HD 200709 
and that deeper searches are required for detection of an optical counterpart.

\begin{table}
\caption{Timing Solution of the transient X-ray pulsar SAX J2103.5+4545
   }
\label{Pri}
\[
\begin{tabular}{ c c }  \hline
Epoch(MJD)  &  51519.8353(8)     \\
Pulse Period (sec) & 358.62171(88)   \\
Derivative of the
Pulse Period (sec $year^{-1}$) & -1.0(9) \\
Orbital Epoch (MJD) & 51519.3(2) \\
P$_{orb}$ (days)    & 12.68(25) \\
a$_{x}$ sin i (lt-sec) (projected semimajor axis) & 72(6) \\
e (eccentricity) & 0.4(2) \\
w (longitude of periastron) & 240(30) \\ \hline

\end{tabular}
\]
\end{table}

{\Large{\bf Figure Caption}}\\

{\bf Fig. 1}~~The total (5 PCU) RXTE/PCA background subtracted X-ray 
light curve.\\
{\bf Fig. 2}~~The pulse profile of SAX J2103.5+4545. The normalized 
count rate is 83 count/sec.\\
{\bf Fig. 3}~~Pulse arrival times of SAX J2103.5+4545 
with respect to the constant pulse period of 358.62171 sec. The quadratic trend 
shows the intrinsic -$\dot P$.\\
{\bf Fig. 4}~~Pulse arrival times of SAX J2103.5+4545 
with respect to the constant pulse period of 358.62171 sec after 
the quadratic trend is removed (above). Solid line denotes the orbit model.
The residuals of pulse arrival times in terms of sigma values  
(below).\\ 
{\bf Fig. 5}~~Folded lightcurve of PCA data at the orbital period. 
The normalized count rate is 83 count/sec. 
Orbital phase 0 corresponds to mean longitude of 90 $^{\circ}$ and 
orbital phase $\sim 0.4$ corresponds to periastron passage.\\
{\bf Fig. 6}~~The light curve of ASM data.\\
{\bf Fig. 7}~~$\chi^{2}$ search of ASM data covering 1997 outburst.\\
{\bf Fig. 8}~~Folded lightcurve of ASM light curve. 
Orbital phase 0 corresponds to mean longitude of 90 $^{\circ}$ and
orbital phase $\sim 0.4$ corresponds to periastron passage.\\

\newpage
\clearpage
\begin{figure}
\plotone{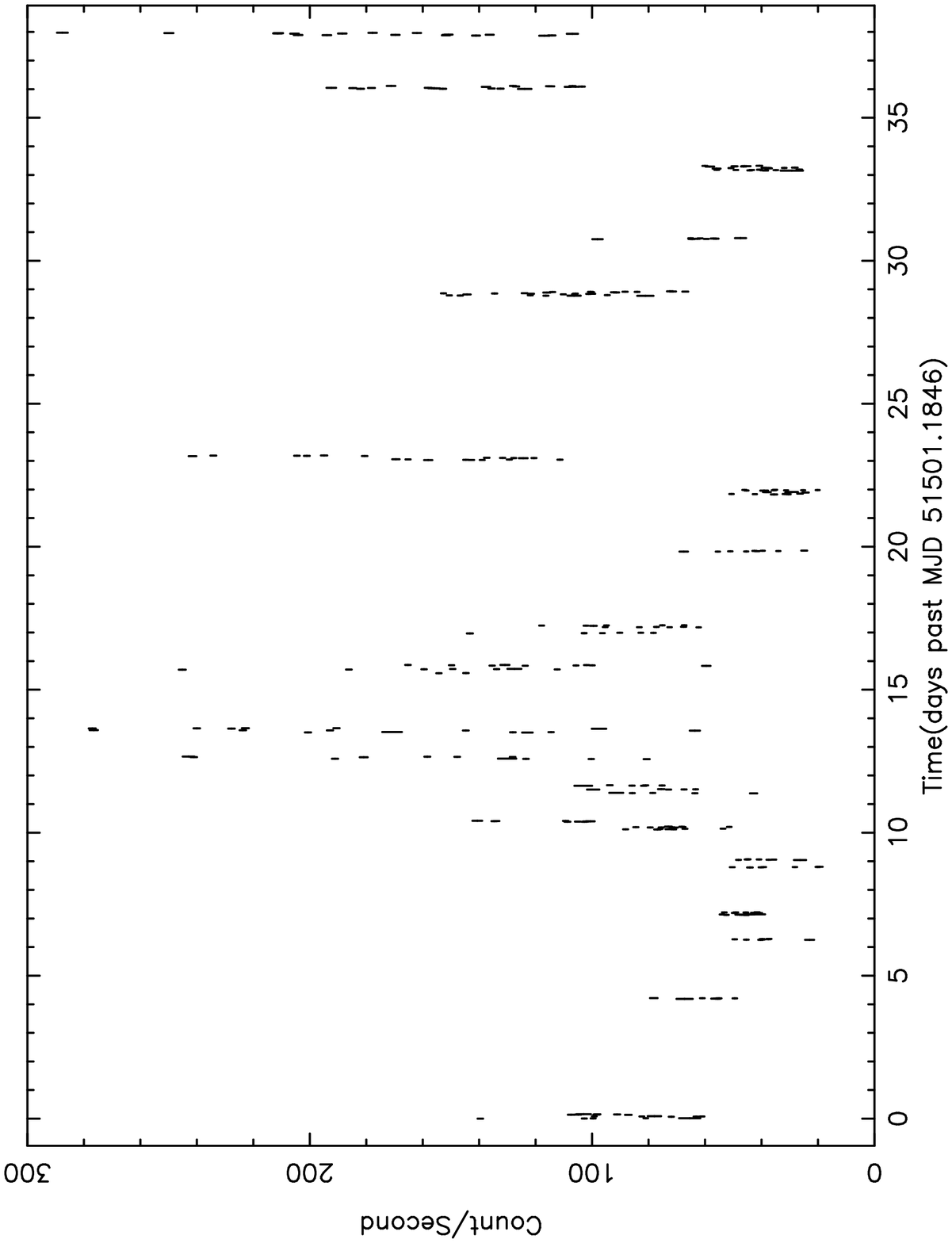}
\end{figure}

\newpage
\clearpage
\begin{figure}
\plotone{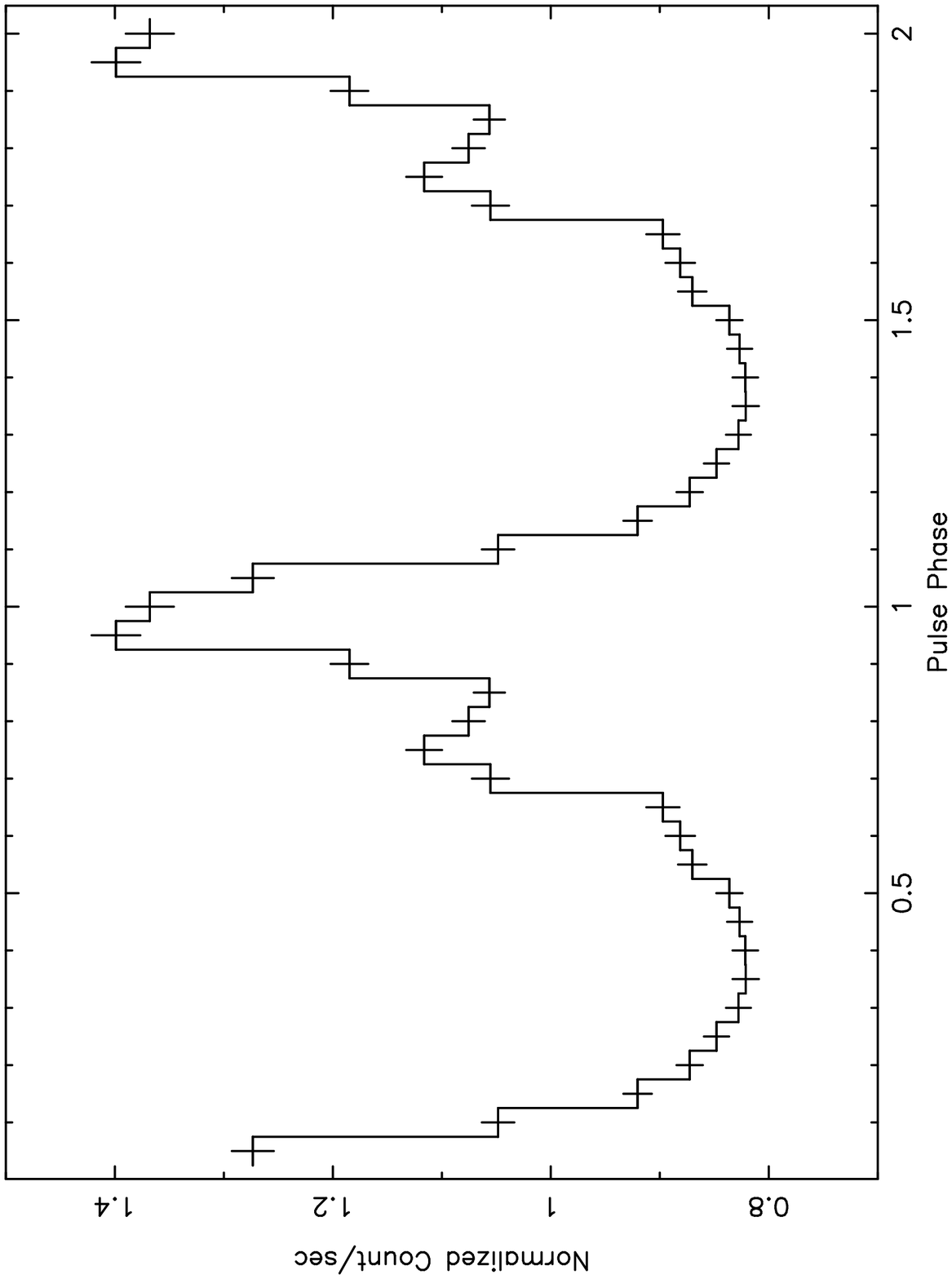}
\end{figure}

\newpage
\clearpage
\begin{figure}
\plotone{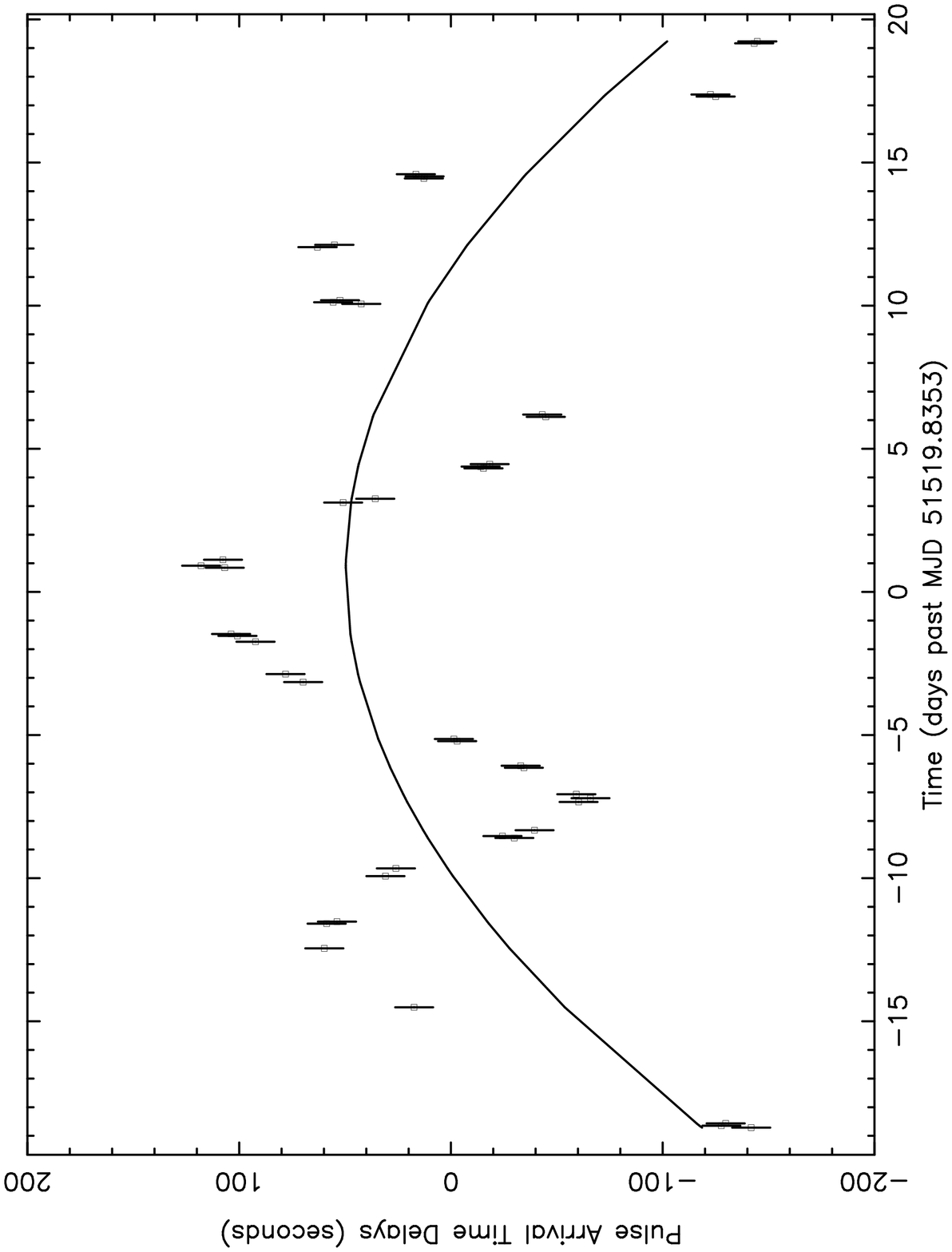}
\end{figure}

\newpage
\clearpage
\begin{figure}
\plotone{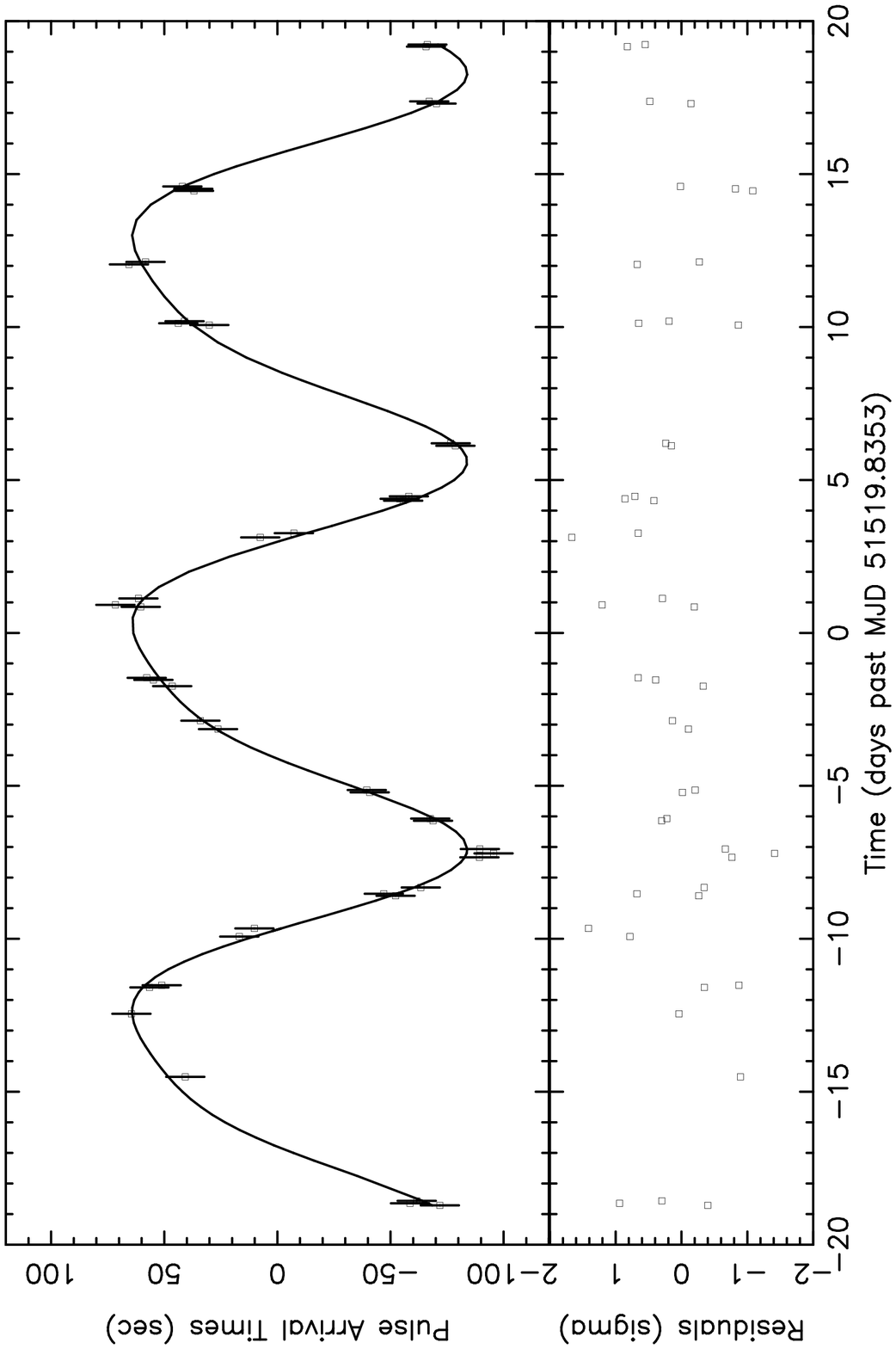}
\end{figure}

\newpage
\clearpage
\begin{figure}
\plotone{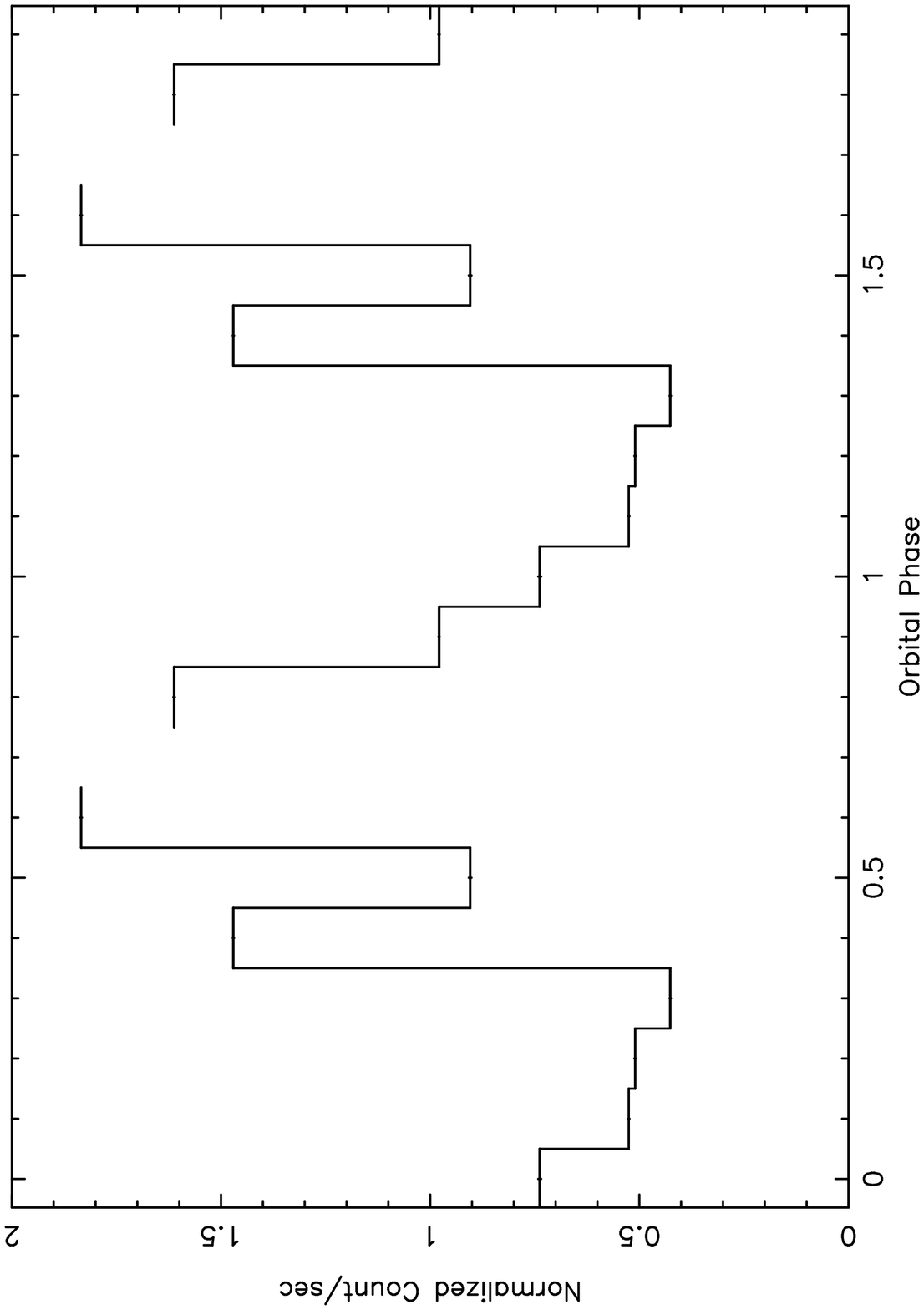}
\end{figure}

\newpage
\clearpage
\begin{figure}
\plotone{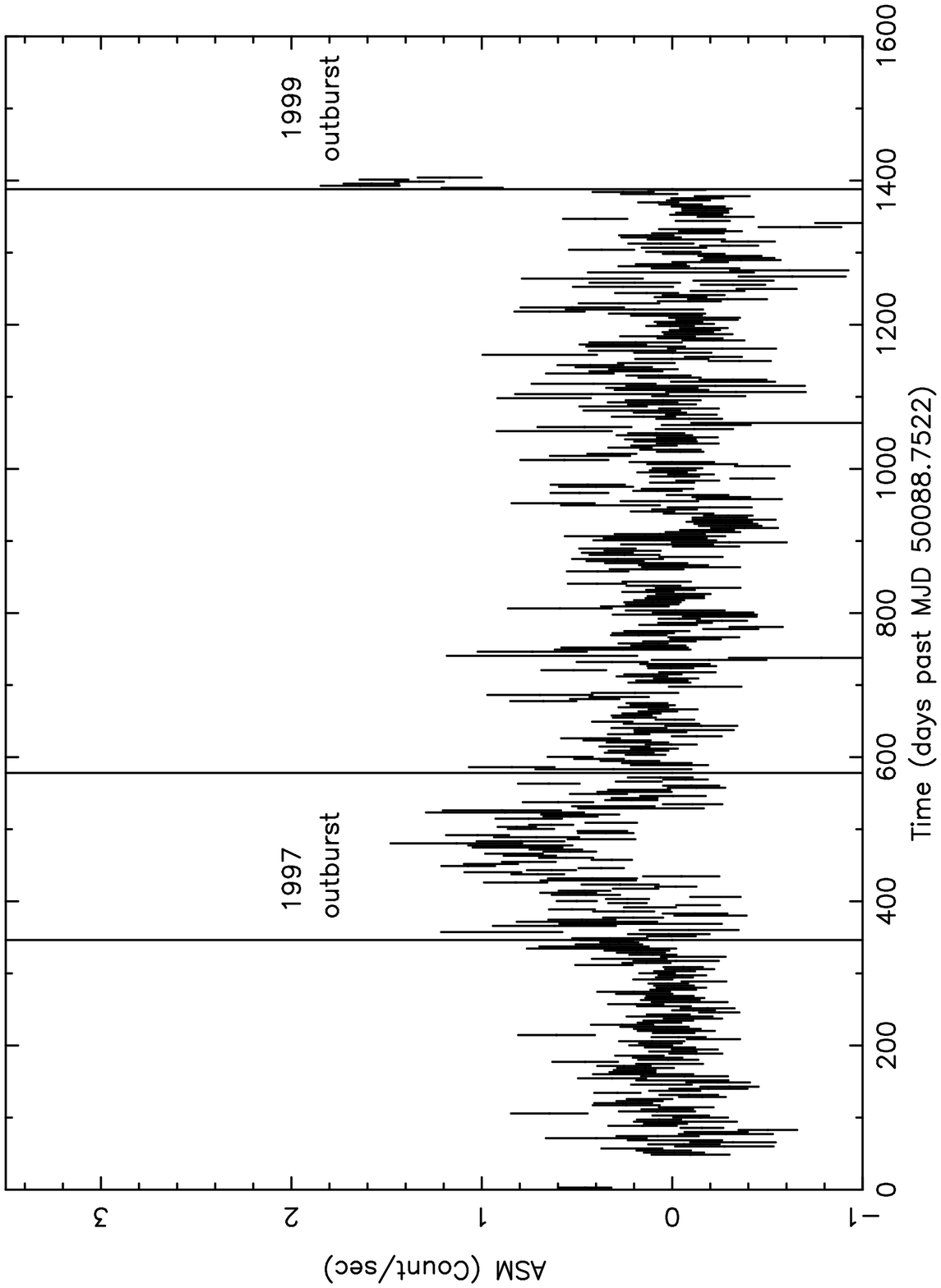}
\end{figure}

\newpage
\clearpage
\begin{figure}
\plotone{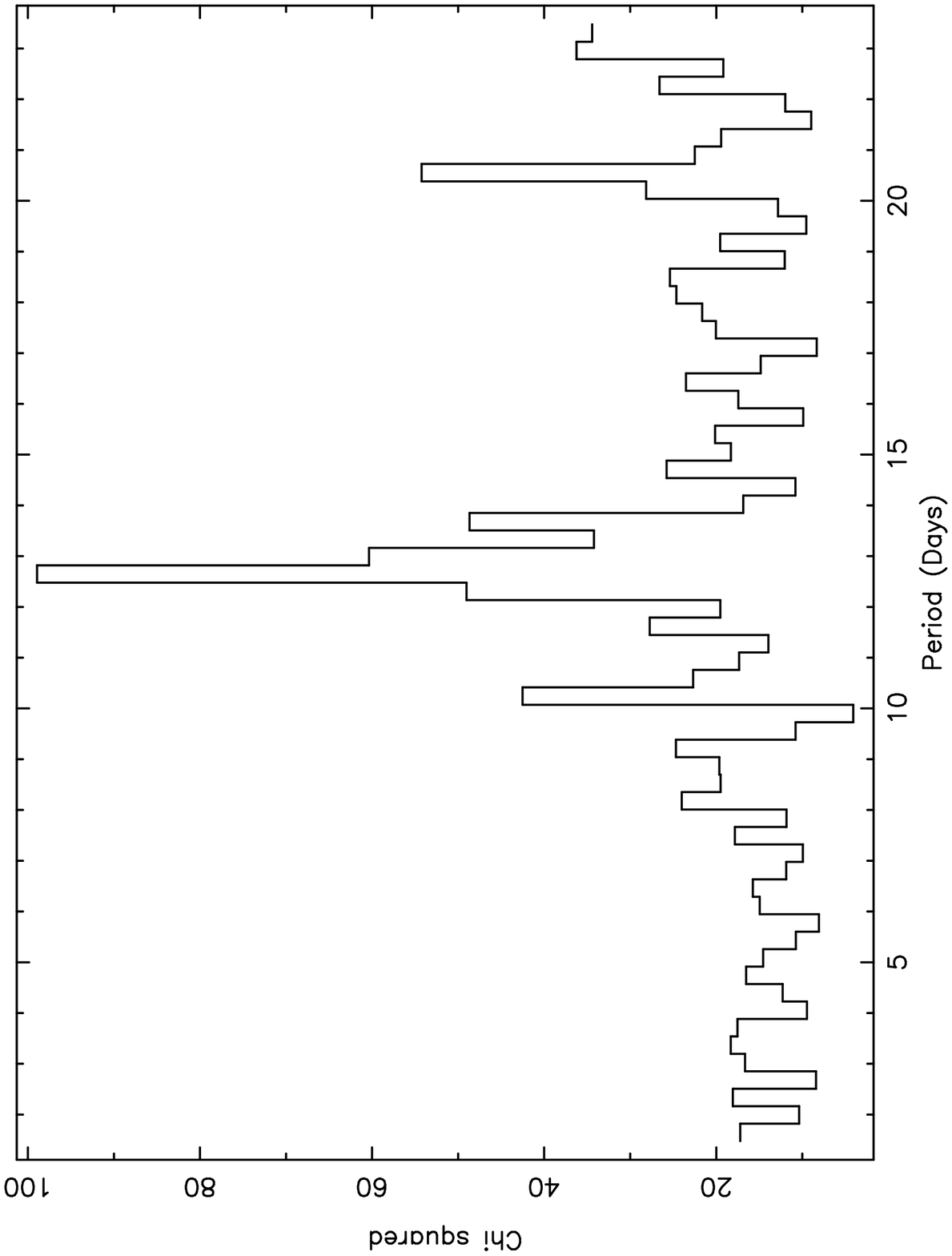}
\end{figure}

\newpage
\clearpage
\begin{figure}
\plotone{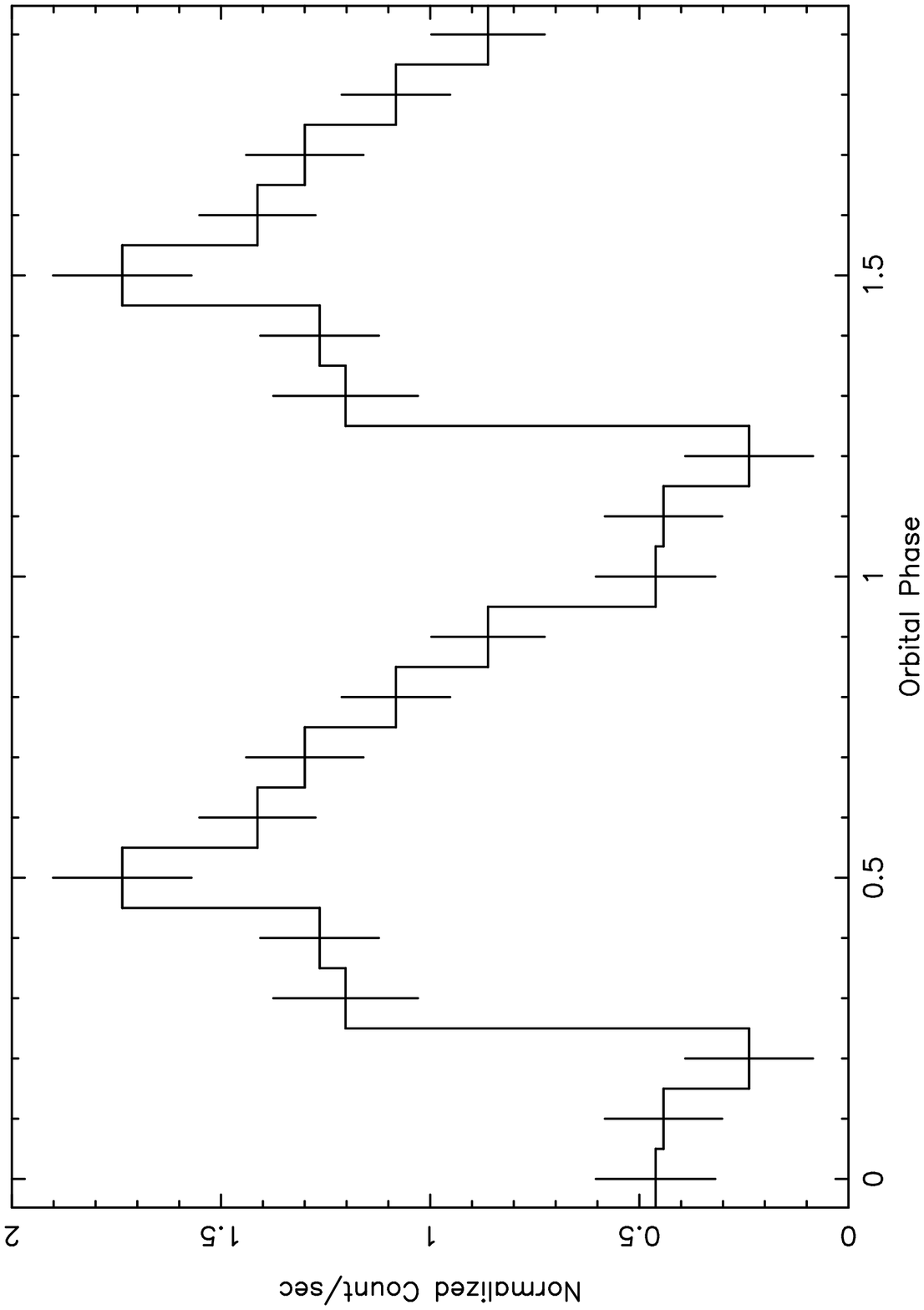}
\end{figure}

\end{document}